\documentstyle[prc,aps]{revtex}
\title{Charge-Dependence of the Nucleon-Nucleon 
Interaction}
\author{G. Q. Li\thanks{Present Address: Department of Physics,
SUNY, Stony Brook, NY 11794} and
R. Machleidt\thanks{Electronic address: machleid@uidaho.edu}}
\address{\it Department of Physics, University of Idaho,
Moscow, ID 83844, U.S.A.}
\date{\today}

\begin{document}

\maketitle

\begin{abstract}
Based upon the Bonn meson-exchange-model for the nucleon-nucleon ($NN$)
interaction, we calculate the charge-independence breaking (CIB)
of the $NN$ interaction due to pion-mass splitting.
Besides the one-pion-exchange (OPE), we take into account
the $2\pi$-exchange model and
contributions from three and four irreducible pion exchanges.
We calculate the CIB differences in the $^1S_0$ effective range parameters
as well as phase shift differences
for partial waves up to total angular momentum $J=4$ and laboratory energies
below 300 MeV.
We find that the CIB effect from OPE dominates in all partial
waves. However, the CIB effects from the $2\pi$ model are
noticable up to D-waves and amount to about 40\% of the OPE CIB-contribution
in some partial waves, at 300 MeV.
The effects from 3$\pi$ and 4$\pi$ contributions 
are  negligible except in $^1S_0$ and $^3P_2$.
\end{abstract}
\pacs{PACS numbers: 24.80.+y, 11.30.Hv, 13.75.Cs, 21.30.Cb}

\twocolumn

\section{Introduction}
It is well known that isospin invariance is not an exact symmetry of
strong interactions. Consequently, nuclear forces have a small but 
measurable charge-dependent component. The equality between 
proton-proton ($pp$) [or neutron-neutron ($nn$)] and neutron-proton ($np$)
nuclear interactions is known as charge independence.
Charge-independence breaking (CIB) is seen most clearly
in the $^1S_0$ nucleon-nucleon
($NN$) scattering lengths. 
The latest empirical values for the singlet scattering length $a$ 
and effective range $r$ are~\cite{MNS90}:
\begin{equation}
\begin{array}{lll}
a^N_{pp}=-17.3\pm 0.4 \mbox{ fm}, &\hspace*{1.0cm}
                                     & r^N_{pp}=2.85\pm 0.04 \mbox{ fm},\\
a^N_{nn}=-18.8\pm 0.3 \mbox{ fm}, && r^N_{nn} = 2.75\pm 0.11 \mbox{ fm},\\
a_{np}=-23.75\pm 0.01 \mbox{ fm}, && r_{np}=2.75\pm 0.05 \mbox{ fm}.
\end{array}
\end{equation}
The values given for $pp$ and $nn$ 
scattering refer to the nuclear part of the interaction
as indicated by the superscript $N$.
Electromagnetic effects have been removed from the experimental
values, which is model dependent. The uncertainties
quoted for $a^N_{pp}$ and $r^N_{pp}$ are mainly due to this model dependence.

It is useful to define the following averages:
\begin{eqnarray}
\bar{a}\equiv \frac12 (a^N_{pp} + a^N_{nn}) & =&  -18.05\pm 0.5 \mbox{ fm},\\
\bar{r}\equiv \frac12 (r^N_{pp} + r^N_{nn}) & =&  2.80\pm 0.12 \mbox{ fm}.
\end{eqnarray}
By definition, CIB is the difference between 
the $np$
values
and 
these averages: 
\begin{eqnarray}
\Delta a_{CIB} \equiv
 \bar{a}
 - 
 a_{np}
 &=& 5.7\pm 0.5 \mbox{ fm},\\
\Delta r_{CIB} \equiv
 \bar{r}
 - 
 r_{np}
 &=& 0.05\pm 0.13 \mbox{ fm}.
\end{eqnarray}
Thus, the $NN$ singlet scattering length shows a clear signature
of CIB in strong interactions.

Charge dependence of $NN$ interactions has been the subject of extensive
investigations, both experimentally and theoretically, for many decades
(for recent reviews, see Refs.~\cite{MNS90,MO95}).
The current understanding is 
that the charge dependence of nuclear forces is due to
a differences in the up and down quark masses and electromagnetic 
interactions. 
On a more phenomenological level, major causes of CIB are:
\begin{itemize}
\item
mass splitting of isovector mesons; particularly, $\pi$ and $\rho$;
\item
irreducible pion-photon exchanges.
\end{itemize}

It has been known for a long time that the difference between the charged and 
neutral pion masses in the one-pion-exchange (OPE) potential  accounts
for about 50\% of $\Delta a_{CIB}$. In Ref. \cite{CM86}, 
charge dependent interactions
were derived for {\it np} and {\it pp} 
scattering, based on a preliminary version of the  
Bonn meson-exchange model~\cite{Mac84}
taking into account the pion mass difference in OPE as well as two-boson
exchanges. With these interactions, about
80\% of the empirical
$\Delta a_{CIB}$ 
could be explained.
Earlier, Ericson and
Miller~\cite{EM83} had obtained a very similar result using the
meson-exchange model of Partovi and Lomon~\cite{PL70}.

The calculations of Refs.~\cite{CM86,EM83} were performed only for 
the singlet scattering length.
However,
it is also of interest to know the charge-dependent effects for intermediate
energies and in partial waves other than $^1S_0$.
Therefore,
it is the main purpose of the present investigation to calculate phase
shift differences between {\it pp} (or $nn$) and {\it np} scattering for states 
with total angular momentum J$\le$ 4 and laboratory incident kinetic energies 
$T_{lab}\le$ 300 MeV.
This paper complements an earlier one on charge-asymmetry of 
the $NN$ interaction~\cite{LM98}.

In Sect.~II, we will discuss
various classes of irreducible meson-exchange diagrams
and calculate their CIB effects---due to pion-mass splitting---on 
$NN$ phase shifts and 
singlet effective range parameters.
Summary and conclusions are given in Sect.~III.

\section{The Bonn Model and Charge-Dependence}

The Bonn meson-exchange model for the $NN$ interaction has been described in
detail in the literature~\cite{MHE87,Mac89,ML93}. 
It is a field-theoretic  model that, apart from the well-known 
one-boson-exchange terms, includes an explicit model for the
$2\pi$-exchange, $\pi\rho$ diagrams, and some further contributions
of $3\pi$- and $4\pi$-exchange. The Bonn model yields
an excellent description of the $NN$ scattering data below
pion production threshold~\cite{ML93} and, thus, provides a reliable basis for
an investigation of the charge dependence of the $NN$ interaction. 
Within the model, charge-dependence is created by the mass difference
between the charge-states of mesons. The Bonn model includes three
isovector mesons, namely, $\pi$, $\rho(770)$, and 
$a_0/\delta(980)$.
We will focus here mainly on the charge-dependent effects 
due to pion mass difference.
Effects due to rho-mass splitting will be discussed briefly at the
end of this section, and
mass splitting of the
$a_0/\delta(980)$ will be ignored since nothing is known.

We use averages for the baryon masses: the average nucleon mass
$M_N=938.919$ MeV and the average $\Delta$-mass $M_\Delta=1232$ MeV.
The pion masses are 
\begin{equation}
m_{\pi^\pm}=139.568~{\rm MeV}, \hspace*{1.5cm} 
m_{\pi^0}=134.974~{\rm MeV}.
\end{equation}
The values are based upon the 1992 Review of Particle
Properties~\cite{PDG92}.

The interaction Lagrangians involving pions are
\begin{eqnarray}
{\cal L}_{\pi NN} & = &  \frac{f_{\pi NN}}{m_{\pi^\pm}}
                           \bar{\psi}  \gamma_\mu \gamma_5 
                           \mbox{\boldmath $\tau$} \psi
                           \cdot \partial^\mu
                           \mbox{\boldmath $\varphi$}_\pi \; ,
\\
{\cal L}_{\pi N\Delta} & = & \frac{f_{\pi N\Delta}}{m_{\pi^\pm}}
                           \bar{\psi} 
                           \mbox{\boldmath $T$} \psi_\mu
                           \cdot \partial^\mu
                           \mbox{\boldmath $\varphi$}_\pi 
                            + \mbox{ H.c.} \; ,
\end{eqnarray}
with $\psi$ the nucleon, $\psi_\mu$ the $\Delta$ (Rarita-Schwinger
spinor), and 
$\mbox{\boldmath $\varphi$}_\pi$ the pion fields.
{\boldmath $\tau$} are the usual Pauli matrices describing
isospin 1/2 and 
{\boldmath $T$}
is the isospin transition operator.
H.c. denotes the Hermitean conjugate.

The Lagrangians are devided by $m_{\pi^\pm}$ to make
the coupling constants $f$ dimensionless.
Following established conventions~\cite{Dum83},
we always use $m_{\pi^\pm}$ as scaling mass.
It may be tempting to use $m_{\pi^0}$ for $\pi^0$ coupling.
Notice, however, that the scaling mass could be anything.
Therefore, it is reasonable to keep the scaling mass constant
within SU(3) multiplets~\cite{Dum83}. This avoids the creation of
unmotivated CIB.

In our investigation of charge-dependent effects on the $NN$ interaction,
we start from a case that may be denoted  the average between $pp$ and
$nn$ scattering. For this case, our model yields $-18.05$ fm for the singlet
scattering length and 2.864 fm for the effective
range, consistent with Eqs.~(2) and (3).
The one-pion-exchange contribution for this average case is depicted in Fig.~1a
and $2\pi$-exchange contributions are shown in Fig.~2a, 3a, and 4a.
Note that, in this case, the proton `p' in Figs.~1a to 4a carries the
average nucleon mass of 938.919 MeV and there are no electromagnetic
interactions; equally well, one may
use a neutron `n' in place of the proton in part (a) of all figures.

To calculate the effects of charge dependence on the $NN$ phase
shifts, we introduce for each  $LSJ$ state the CIB
phase shift difference $\Delta\delta^{LSJ}_{CIB}(T_{lab})$, defined by
\begin{equation}
\Delta\delta^{LSJ}_{CIB}(T_{lab}) \equiv
 \delta^{LSJ}_{np}(T_{lab})
-
 \bar{\delta}^{LSJ}(T_{lab})
\end{equation}
where $\bar{\delta}^{LSJ}$ denotes the average of the $pp$ and $nn$
phase shifts
which, as discussed, is calculated by taking 
the diagrams Figs.~1a to 4a into account
(besides the other diagrams
involved in the Bonn model) 
with average nucleon mass and all electromagnetic interactions
switched off.
The phase shift $\delta^{LSJ}_{np}$
is the $np$ one to be calculated below.
Similarly, we define the CIB mixing parameter difference 
$\Delta\epsilon^J_{CIB}$,
\begin{equation}
\Delta\epsilon^J_{CIB}(T_{lab}) \equiv
 \epsilon ^J_{np}(T_{lab})
-
 \bar{\epsilon}^J (T_{lab})
 \: .
\end{equation}

The charge-dependence generated by the model under consideration
is now `switched on' step by step: 

\begin{enumerate}

\item
{\bf One-pion-exchange} (OPE), Fig.~1: 
The CIB effect is created by replacing the diagram of Fig.~1a
by the two diagrams of Fig.~1b. 
Note that one-meson-exchange contributions are roughly
proportional to 1/$m_\alpha^2$ (with $m_\alpha$ the meson mass)
because this is approximately the momentum-space one-meson propagator
for very low momentum transfer.
Thus, since the $\pi^0$ has a smaller mass than the $\pi^\pm$,
$\pi^0$ exchange is stronger than $\pi^\pm$ exchange.
For this reason, OPE is stronger in $pp$ as
compared to $np$.
Since OPE is repulsive in $^1S_0$, this phase shift becomes
more attractive (i.~e., larger) when going from $pp$ to $np$,
resulting in a positive $\Delta \delta$; cf.\ column `OPE'
in Table~I and dashed curve in Fig.~5.
Consistent with this
is the well-known fact that OPE takes care of about 
50\% of $\Delta a_{CIB}$. 
In the other partial waves, the sign of $\Delta \delta$ due to OPE
depends on if OPE is repulsive or attractive (e.~g., it is repulsive in
$^3P_1$ and attractive in $^3P_0$ and $^3P_2$).
Due to the small mass of the pion, OPE is a sizable
contribution in all partial waves including higher partial waves; 
and due to the pion's relatively large mass splitting (3.4\%),
OPE creates relatively large charge-dependent effects 
in all partial waves (Fig.~5 and Table~I).

\item
{\bf $2\pi$-exchange with $NN$ intermediate
states} ($2\pi NN$), Fig~2: Notice first that
only non-iterative (irreducible) diagrams are to be considered, since
the iterative ones  
are generated by the scattering equation from
OPE. We mention here that, in our approach which is based upon time-ordered
perturbation theory,
we always take all time-orderings into account (except for those that imply
 anti-baryons in intermediate states);
however, to save space, we display only a few characteristic
time-orderings in Fig.~2 (this is also true for all diagrams shown
in Figs.~3 and 4; to get an impression of the total number
of time-ordered diagrams, see Fig.~20 of Ref.~\cite{MHE87}). 
The CIB effect is obtained by replacing the diagrams Fig.~2a
($pp$/$nn$ scattering) by those of Fig.~2b ($np$ scattering).
For a good understanding of CIB effects, it is important to
distinguish between box (here: stretched box) and crossed
box diagrams.
Concerning the effect from stretched box diagrams,
one replaces the left diagram of Fig.~2a 
by the four stretched box diagrams
of Fig.~2b (and similarly for the other stretched box
time-orderings not shown).
Notice now that in the former diagram two $\pi^0$
are exchanged making this a `strong' diagram,
while the latter four diagrams together with their
isospin factors result in a weaker contribution.
Since 2$\pi$ exchange is, in general, attractive, there
is a loss of attraction when going from $pp$ to $np$
(equivalent to a reduction of the phase shift).
This qualitative estimate is clearly confirmed
by the quantitative results displayed in column `$2\pi NN-S$'
of Table~II.
The CIB effect that stems from crossed box diagrams
is obtained by replacing the two crossed boxes of Fig.~2a
by the three crossed boxes of Fig.~2b.
Typically, this effect (column `$2\pi NN-X$' of Table~II)
is of opposite
sign as compared to the corresponding (stretched) 
box effect, in most partial waves.
The total CIB effect from all diagrams of Fig.~2 is displayed
by the dashed curve in Fig.~6.

\item
{\bf $2\pi$-exchange with N$\Delta$ intermediate states}
($2\pi N\Delta$),
Fig.~3:
Notice again, that every box stands for all possible 
time-orderings of the box type (diagrams 1 to 6 of Fig.~20
of Ref.~\cite{MHE87})
and every crossed box for all possible time-orderings
of the crossed box type (diagrams 7 to 12 of Fig.~20 of Ref.~\cite{MHE87}).
Thus, the total number of diagrams which Fig.~3a stands for is 24
while Fig.~3b stands for 48 diagrams, which are all explicitly
taken into account in our calculations.
Replacement of the boxes Fig.~3a by the boxes Fig.~3b
causes an increase in the strength of these diagrams
which, since these are attractive diagrams, causes an increase in
attraction. Column `$2\pi N\Delta-B$' in Table~II clearly confirms this.
For the crossed boxes one gets typically the opposite
effect (column `$2\pi N\Delta-X$' of Table~II).
This partial cancelation of the effects from the two groups
of diagrams is also demonstrated
in Fig.~6 where the dash-dot curve represents the
effect from the box diagrams while the dash-triple-dot
curve is from the crossed ones. Notice that the cancelation
is almost perfect in $^1S_0$ and $^3P_2$ even though the individual
contributions are rather large.

\item
{\bf $2\pi$-exchange with $\Delta\Delta$ intermediate states}
($2\pi\Delta\Delta$),
Fig.~4:
The replacement of the diagrams of Fig.~4a by Fig.~4b
shows the by now familiar characteristic: 
opposite effects from box and crossed box diagrams
(column `$2\pi\Delta\Delta-B$'
and `$2\pi\Delta\Delta-X$'
of Table~II).
This results in
large
cancelations between effects which,
due to the
short-range nature of this class of diagrams, 
are individually already rather small.
This explains why the CIB effect from the diagrams of Fig.~4
is negligible in most partial waves (dotted curve in Fig.~6).

This finishes the discussion of all contributions of the
$2\pi$ type. In summary, one can say that the total CIB effect
from $2\pi$ (dash-dot curve in Fig.~5)
is quite noticable up to the $D$-state. 
In $^1S_0$, $^3P_0$, $^3P_1$, and $^1D_2$, the CIB effect from $2\pi$ is 
20-50\% of the one from OPE, at 300 MeV.
However, for low energies (except in $^1S_0$)
as well as in higher partial waves, the CIB $2\pi$ effect is negligible.

\item
{\bf  $\pi\rho$-exchanges.} This group is, in principal, as comprehensive
as the $2\pi$-exchanges discussed above. Graphically, 
the $\pi\rho$ diagrams can be obtained
by replacing in each diagram of Figs.~2--4, one of the two pions by a
$\rho$-meson of the same charge-state (because of this simple
analogy, we do not show the $\pi\rho$ diagrams explicitly here).
Concerning the $\pi\rho$ diagrams
with $\Delta$ intermediate
states a comment is in place. In the Bonn model~\cite{MHE87},
the crossed $\pi\rho$ diagrams with $N\Delta$ 
and $\Delta\Delta$ intermediate
states 
are included in terms of an approximation.
It is assumed that they differ from the corresponding box diagrams
only by the isospin factor. Thus, the $\pi\rho$
box diagrams with $N\Delta$ and
$\Delta\Delta$ intermediate states are multiplied by an
isospin factor that is equal to the sum of the isospin factors
 for box and crossed box.
In this approximation, these diagrams do not
generate any
CIB effects due to pion-mass splitting.
Since these
diagrams are of very short range, their CIB effect
may be negligible, anyhow.
The only class of $\pi\rho$ diagrams which we include in our
calculation of CIB effects is the one that corresponds to Fig.~2,
with one pion in each diagram replaced by a $\rho$ meson.
Its contribution to CIB (column '$\pi\rho$' of Table~I and 
dash-triple-dot curve of Fig.~5) is generally small, and
typically opposite to the one from $2\pi$,
in most states.

\item
{\bf Further $3\pi$ and $4\pi$ contributions} ($\pi\sigma+\pi\omega$).
The Bonn potential also includes some $3\pi$-exchanges that can be
approximated in terms of $\pi\sigma$ diagrams and $4\pi$-exchanges
of  $\pi\omega$ type.
It was found in Ref.~\cite{MHE87} that
the sum of these contributions is small. 
These diagrams have $NN$ intermediate states---similar to Fig.~2,
but with one of the two exchanged pions replaced by an isospin-zero boson;
thus, the isospin factors are different from the ones of Fig.~2 and,
in fact, like the ones of Fig.~1. 
Another way of creating these diagrams is to combine the
diagrams of Fig.~1 with a sigma or an omega in an
irreducible way, i.~e., by forming a stretched box or
crossed box diagram. These diagrams carry the same isospin factors
as OPE.
Since this class of diagrams is part of the Bonn model,
we include these diagrams in our CIB consideration.
The CIB effect from this class is very small, except in
$^1S_0$, $^3P_1$, and $^3P_2$ (Column `$\pi\sigma + \pi\omega$'
of Table~I and dotted curve in Fig.~5).
This effect has always the same sign as the effect from OPE, 
but it is substantially
smaller. 
The reason for the OPE character of this contribution is that
$\pi\sigma$ prevails over $\pi\omega$ and, thus, determines the
character of this
contribution. Since sigma-exchange is negative and since, futhermore,
the propagator in between the $\pi$ and the $\sigma$ exchange is also
negative, the overall sign of the $\pi\sigma$ exchange is the same as OPE.
Thus, it is like a weak, short-ranged OPE.
\end{enumerate}

This finishes our detailed presentation of the diagrams and their
CIB effects included
in our calculation. The sum of all CIB effects on phase shifts is given in 
the last column of Table~I and plotted by the solid curve in Fig.~5.
Notice that the difference between the solid curve
and the dashed curve (OPE) in Fig.~5 represents the sum of
all effects beyond OPE.
Thus, it is clearly seen that OPE dominates the CIB effect in all
partial waves, even though there are substantial contributions besides
OPE in some states, notably $^1S_0$, $^3P_1$, and $^1D_2$.

Finally, in Table III and IV, we also give the CIB contributions
to the $^1S_0$ scattering length and effective range.
Note that the relationship between the CIB potential and the
corresponding change of the
scattering length, $\Delta a_{CIB}$, is highly non-linear. 
As discussed in Refs.~\cite{EM83,CM86}, when the scattering
length changes from $a_1$ to $a_2$ due to a CIB potential
$\Delta V=V_1 - V_2$, the relationship is
\begin{equation}
\frac{1}{a_2} - \frac{1}{a_1} = M_N
 \int_0^\infty \Delta V u_1 u_2 dr
\end {equation}
or
\begin{equation}
a_1 - a_2 = a_1 a_2 M_N
 \int_0^\infty \Delta V u_1 u_2 dr \; ,
\end {equation}
with $u_1$ and $u_2$ the zero-energy $^1S_0$ wave functions
normalized such that $u(r\rightarrow \infty ) \longrightarrow
(1-r/a)$. Thus, the perturbation expansion concerns the
invers scattering length.
As clearly evident from Eq.~(12),
the change of the scattering length 
depends on the ``starting value'' $a_1$
to which the effect is added.
In our calculations, CIB effects are generated step by step,
which implies that the starting value $a_1$ is different
for each CIB effect.
This distorts the relative size of different CIB contributions
to the scattering length difference.
To make the relative comparison meaningful,
we have rescaled our results for $\Delta a_{CIB}$
according to a prescription given by Ericson and Miller~\cite{EM83},
which goes as follows.
Assume the ``starting value'' for the scattering length is $a_1$
and a certain CIB effect brings it up to $a_2$. 
Then, the resulting scattering length difference $(a_1-a_2)$ is rescaled by
\begin{equation}
\Delta a = (a_1 - a_2) \frac{\bar{a} a_{np}}{a_1 a_2}
\end{equation}
with $\bar{a}=-18.05$ fm and $a_{np}=-23.75$ fm.
This will make $\Delta a$ independent of the choice
for $a_1$.
The numbers given in Table III and IV for $\Delta a_{CIB}$
are all rescaled according to Eq.~(13).

We obtain a total $\Delta a_{CIB}$ of 4.65 fm which is about
80\% of the empirical value of 5.7 fm [Eq.~(4)].
For $\Delta r_{CIB}$ we find a total of 0.115 fm from all effects,
consistent with the empirical value [Eq.~(5)].
Even though our total result for $\Delta a_{CIB}$ is very similar
to the earlier calculation of Ref.~\cite{CM86},
there are some differences in the details.
For the total effect from $2\pi$-exchange we obtain in the
present calculations $\Delta a_{CIB}=0.36$ fm while in
Ref.~\cite{CM86} 0.85 fm was reported.
This is due to differences in the interpretation of the scaling
mass that occurs in the $N\Delta\pi$ Lagrangian, Eq.~(8).
While in the present calculations we always use $m_{\pi^\pm}$
[see discussion below Eq.~(8)],
in the earlier calculations of Ref.~\cite{CM86} $m_{\pi^0}$
was used for $\pi^0$ coupling and $m_{\pi^\pm}$ 
for $\pi^\pm$ coupling.
The latter convention introduces a strong charge-dependence
of the effective $N\Delta\pi$ coupling strength,
which enhances the CIB effects from all diagrams involving
$\Delta$-isobars. 
In principal, there is discretion in how to deal with the scaling
mass in Eq.~(8). However, in the present calculations, we decided to
follow the established convention~\cite{Dum83}. As a result,
the effect from $2\pi$-exchange is smaller than in the
earlier calculation of Ref.~\cite{CM86}.

There is also a small difference in the $\Delta a_{CIB}$
contribution
from non-iterative $\pi\sigma$ and $\pi\omega$ exchanges
for which we obtain 1.4 fm while Ref~\cite{CM86} reported
1.2 fm.
This discrepancy is due to the fact that in
Ref.~\cite{CM86} a preliminary version of the Bonn Full Model
\cite{Mac84} was used, while here we applied the final version
\cite{MHE87} in which the strength of the $\pi\sigma$
contribution is slightly larger, which explains the difference.

We note that the CIB effect depends on the $\pi NN$ coupling
constant. In the present calculations, we follow the Bonn 
model~\cite{MHE87}: we assume charge-independence of the
coupling constant and use 
$g^2_\pi/4\pi = 14.4$~\cite{foot}.
In recent years, there has been some controversy about the
precise value of the $\pi NN$ coupling constant.
Unfortunately, the problem is far from being settled.
Based upon $NN$ phase shift analysis,
the Nijmegen group~\cite{Sto93} advocates 
the `small' charge-independent
value
$g^2_\pi/4\pi = 13.5(1)$,
while a very recent determination
by the Uppsala group~\cite{Rah98} based upon high
precision $np$ charge-exchange data at 162 MeV resulted
in the `large' value
$g^2_{\pi^\pm}/4\pi = 14.52(26)$.
Other recent determinations 
are in-between the two extremes:
The VPI group~\cite{Arn94} quotes
$g^2_\pi/4\pi = 13.77(15)$ from $\pi N$ and $NN$
analysis with no evidence for charge-dependence.
Bugg and Machleidt~\cite{BM95} obtain
$g^2_{\pi^\pm}/4\pi = 13.69(39)$ and
$g^2_{\pi^0}/4\pi = 13.94(24)$
from the analysis of $NN$ elastic data between
210 and 800 MeV.
Because of this large uncertainty in the $\pi NN$
coupling constant, it might be of interest to know
what the CIB effects are like when a value is used
that deviates substantially from our choice.
For that reason, we have repeated our CIB calculations
with the smaller values 
$g^2_\pi/4\pi = 14.0$
and 13.6.
It turns out that the total CIB effect on phase shifts
(last column of Table I) as well as the effect on
the effective range parameters (Table III) scales linearly with the 
$\pi NN$ coupling constant, to a good approximation.
To be precise: multiplying the total phase shift
differences in Table I or the effective range changes
in Table III with 13.6/14.4 reproduces
within $\pm 2$\% the exact results from a CIB calculation
that employes
$g^2_\pi/4\pi = 13.6$.

As last item in our study, we have also investigated the effect
of rho-mass splitting on the $^1S_0$ effective range parameters.
Unfortunately, the evidence for rho-mass splitting is very
uncertain, with the Particle Data Group~\cite{PDG92} reporting
$m_{\rho^0} - m_{\rho^\pm} = 0.3 \pm 2.2$ MeV.
Consistent with this, we assumed in our exploratory study
$m_{\rho^0} = 769$ MeV and
$m_{\rho^\pm} = 768$ MeV, i.~e., a splitting of 1 MeV.
With this, we find 
$\Delta a_{CIB} = -0.29$ fm 
from  
one-rho-exchange, and 
$\Delta a_{CIB} = 0.28$ fm 
from the non-iterative $\pi\rho$ diagrams with $NN$ intermediate
states. Thus, individual effects are small and, in addition,
there are substantial cancellations between
the two classes of diagrams that contribute.
The net result is a vanishing effect.
Thus, even if the rho-mass splitting
will be better determined in the future and may turn out to be larger than
our assumption,
it will never be a great source of CIB.

\section{Summary and Conclusions}

Based upon the Bonn meson-exchange model for the $NN$ interaction,
we have calulated the CIB effects due to pion-mass splitting
on the singlet effective range parameters and
on the phase shifts of $NN$ scattering
for partial waves of total angular momentum $J\leq 4$ and
laboratory energies below 300 MeV.
This investigation complements our recent paper on charge-asymmetry
of the $NN$ interaction~\cite{LM98}.

The overall results may be characterized as follows.

The largest phase shift differences occur in the $^1S_0$ state
where they are most noticable at
low energy; e.~g., at 1 MeV, the difference is 4.36$^0$, indicating that the
{\it np} nuclear force is more attractive than the {\it pp} nuclear force.
The $^1S_0$ phase shift difference decreases with increasing energy
and is about 0.6$^0$ at 300 MeV.
The major part of the phase shift difference comes from OPE. 
CIB contributions from two-meson exchange diagrams can be large,
but there are typically cancelations between the effects from different classes
of diagrams of the two-meson type.

The CIB effect on the phase shifts of
$P$ and higher partial waves is generally small.
The most significant difference is found in the $^3P_0$
state around 50 MeV where the difference is almost 1 degree. 
In $P$-waves, the difference is roughly constant above 25 MeV:
it is 
0.95-0.65$^0$
in $^3P_0$, 
about 0.35$^0$
in $^3P_1$, 
and 
around 0.2$^0$
in $^3P_2$. 
In all other partial waves, it is in the order of 0.1$^0$ or less.
Again, the main effect comes from OPE, however, 
in $^3P_0$, $^3P_1$, and $^1D_2$ at 300 MeV,
the effect from 
the $2\pi$ model is in the range of 20 - 50\% of the one from OPE.

The fact that the
magnitudes of the phase shift differences in all partial waves,
except  
$^1S_0$, are small, makes it difficult to verify experimentally 
the charge dependent effects in $P$ and higher partial waves. However, since
the phase shifts in these states themselves are small, the relative magnitudes
of the phase shift differences are not negligible and could have a noticable
effect on some sensitive observables such as the analyzing power ($A_y$) in 
nucleon-deuteron ($nd$)
scattering~\cite{WG92} since this reaction blows up effects from
triplet $P$ waves~\cite{Sla91}. Our microscopic predictions
are, however, 
substantially smaller than what is needed
to solve the $nd$ $A_y$ puzzle~\cite{WG92,HF98}.

As mentioned in the Introduction, another CIB contribution to the nuclear
force is irreducible pion-photon ($\pi\gamma$) exchange.
Traditionally, it was believed that this contribution would take care
of the remaining 20\% of $\Delta a_{CIB}$~\cite{EM83,MO95}.
However, a recently derived $\pi\gamma$ potential based upon chiral
perturbation theory~\cite{Kol98} {\it decreases} $\Delta a_{CIB}$
by about 0.6 fm, making the discrepancy even larger.
Thus, we are faced with the fact that about 20 - 30\%
of the charge-dependence of the singlet scattering length is
not explained.

In recent years, nuclear physicists have become increasingly concerned
with chiral symmetry which is an approximate symmetry of QCD
in the light-quark sector. In the light of these new views,
the $NN$ interaction should have a clear relationship
with chiral symmetry. The Bonn model that our investigation
in based upon is, by construction, not a consistently
chiral model. Chiral models for the $NN$ interaction and,
in particular, chiral models for the $2\pi$ exchange have
recently been constructed by various 
groups~\cite{ORK96,RR97,KBW97}.
However, most of these models are applicable only for the peripheral
partial waves of $NN$ scattering and not for $S$, $P$, or $D$ waves;
and if there are predictions for lower partial waves, they are only
of qualitative nature. The CIB effects in $S$ and $P$ waves and,
particularly, for the singlet scattering
length are very subtle and, therefore, require a
quantitative model. Thus, current chiral models for the
$2\pi$ exchange are not (yet) suitable for reliable
calculations of CIB. One may then raise an interesting
question: What has to be changed in the Bonn model
to make it chiral? This question can be answered precisely.
The diagrams in Figs.~2 (a) and (b) of Ref.~\cite{ORK96}
have to be added to the Bonn model; that is essentially all.
These diagrams include the 
Weinberg-Tomozawa 
$\pi\pi NN$
vertex which is a characteristic ingredient of any nonlinear
realization of chiral symmetry. However, it has been found independently
by different groups~\cite{ORK96,RR97,KBW97} that the $2\pi$ exchange
diagrams which include the
Weinberg-Tomozawa vertex make a very small, essentially negligible,
contribution to the $NN$ interaction. One may then expect that the CIB
caused by these diagrams is also very small~\cite{K98}.
Thus, there are reasons to believe that the results of this study
may be of broader relevance than what the (formally) non-chiral
character of our model suggests.
Of course, the final and reliable answer of the question under
consideration can only come from a `perfect' and quantitative
chiral model for the $NN$ interaction that is applicable
also in $S$ waves and for the calculation of scattering lengths.
In view of the problems raised concerning scattering length
calculations with chiral models~\cite{Sca97,KSW96}
and in view of the continuing general controversy concerning
cutoff {\it versus} dimensional
regularization, it will take many years until a reliable
calculation of this kind can be done.
Thus, for the time being, it may be comforting to have
at least our present results.

\vskip 1cm
This work was supported in part by the U.S. 
National Science Foundation under Grant No.~PHY-9603097 and 
by the Idaho State Board of Education.

\pagebreak

\pagebreak

\onecolumn

\begin{table}
\caption{CIB phase differences (in degrees)
 as defined in Eqs.~(9) and (10) and explained in the text.}
\begin{tabular}{rrrrrr}
T$_{lab}$ (MeV)& OPE & $2\pi$  & $\pi\rho$ & $\pi\sigma+\pi\omega$ & Total  \\
 \hline 
 \hline 
 \\ \multicolumn{6}{c}{$^1S_0$}  \\
     1     &    3.051   &    0.319   &   -0.361   &    1.355   &    4.364   \\
     5     &    1.767   &    0.154   &   -0.193   &    0.719   &    2.446   \\
    10     &    1.364   &    0.103   &   -0.147   &    0.545   &    1.864   \\
    25     &    0.944   &    0.047   &   -0.107   &    0.391   &    1.275   \\
    50     &    0.712   &    0.009   &   -0.090   &    0.318   &    0.950   \\
   100     &    0.563   &   -0.031   &   -0.083   &    0.276   &    0.725   \\
   150     &    0.519   &   -0.061   &   -0.086   &    0.267   &    0.638   \\
   200     &    0.509   &   -0.091   &   -0.093   &    0.269   &    0.595   \\
   300     &    0.544   &   -0.163   &   -0.121   &    0.300   &    0.559   \\
 \hline 
 \\ \multicolumn{6}{c}{$^3P_0$}  \\
     1     &   -0.032   &    0.000   &    0.000   &    0.000   &   -0.032   \\
     5     &   -0.246   &   -0.003   &    0.000   &   -0.001   &   -0.250   \\
    10     &   -0.482   &   -0.009   &    0.001   &   -0.002   &   -0.492   \\
    25     &   -0.827   &   -0.029   &    0.002   &   -0.005   &   -0.858   \\
    50     &   -0.902   &   -0.053   &    0.005   &   -0.009   &   -0.960   \\
   100     &   -0.786   &   -0.078   &    0.008   &   -0.012   &   -0.869   \\
   150     &   -0.685   &   -0.091   &    0.009   &   -0.011   &   -0.778   \\
   200     &   -0.618   &   -0.101   &    0.011   &   -0.009   &   -0.717   \\
   300     &   -0.540   &   -0.118   &    0.012   &   -0.002   &   -0.648   \\
 \hline 
 \\ \multicolumn{6}{c}{$^3P_1$}  \\
     1     &    0.017   &    0.000   &    0.000   &    0.000   &    0.017   \\
     5     &    0.116   &    0.001   &    0.000   &    0.000   &    0.117   \\
    10     &    0.203   &    0.002   &    0.000   &    0.001   &    0.206   \\
    25     &    0.313   &    0.007   &   -0.001   &    0.003   &    0.322   \\
    50     &    0.346   &    0.016   &   -0.004   &    0.008   &    0.366   \\
   100     &    0.323   &    0.032   &   -0.009   &    0.018   &    0.364   \\
   150     &    0.289   &    0.046   &   -0.014   &    0.028   &    0.349   \\
   200     &    0.259   &    0.058   &   -0.020   &    0.037   &    0.335   \\
   300     &    0.213   &    0.083   &   -0.032   &    0.057   &    0.321   \\
 \hline 
 \\ \multicolumn{6}{c}{$^1D_2$}  \\
     5     &   -0.009   &    0.000   &    0.000   &    0.000   &   -0.009   \\
    10     &   -0.025   &    0.000   &    0.000   &    0.000   &   -0.025   \\
    25     &   -0.052   &    0.001   &    0.000   &    0.000   &   -0.050   \\
    50     &   -0.045   &    0.005   &    0.000   &    0.000   &   -0.040   \\
   100     &    0.003   &    0.015   &   -0.001   &    0.001   &    0.018   \\
   150     &    0.044   &    0.025   &   -0.001   &    0.001   &    0.069   \\
   200     &    0.071   &    0.034   &   -0.002   &    0.002   &    0.106   \\
   300     &    0.100   &    0.047   &   -0.002   &    0.002   &    0.147   \\
 \hline 
 \\ \multicolumn{6}{c}{$^3P_2$}  \\
     5     &   -0.010   &    0.000   &    0.000   &    0.000   &   -0.011   \\
    10     &   -0.030   &    0.000   &    0.000   &   -0.001   &   -0.032   \\
    25     &   -0.096   &   -0.001   &   -0.001   &   -0.003   &   -0.101   \\
    50     &   -0.172   &   -0.003   &   -0.002   &   -0.007   &   -0.184   \\
   100     &   -0.224   &   -0.007   &   -0.005   &   -0.014   &   -0.250   \\
   150     &   -0.224   &   -0.007   &   -0.007   &   -0.019   &   -0.258   \\
   200     &   -0.210   &   -0.007   &   -0.008   &   -0.022   &   -0.248   \\
   300     &   -0.183   &   -0.004   &   -0.010   &   -0.022   &   -0.219   \\
 \hline 
 \\ \multicolumn{6}{c}{$^3F_2$}  \\
    10     &   -0.004   &    0.000   &    0.000   &    0.000   &   -0.004   \\
    25     &   -0.020   &    0.000   &    0.000   &    0.000   &   -0.020   \\
    50     &   -0.045   &    0.000   &    0.000   &    0.000   &   -0.045   \\
   100     &   -0.070   &   -0.001   &    0.000   &    0.000   &   -0.071   \\
   150     &   -0.084   &   -0.001   &    0.000   &    0.000   &   -0.084   \\
   200     &   -0.093   &   -0.001   &    0.000   &    0.000   &   -0.093   \\
   300     &   -0.102   &   -0.001   &   -0.001   &    0.001   &   -0.102   \\
 \hline 
 \\ \multicolumn{6}{c}{$\epsilon_2$}  \\
     5     &    0.012   &    0.000   &    0.000   &    0.000   &    0.012   \\
    10     &    0.036   &    0.000   &    0.000   &    0.000   &    0.036   \\
    25     &    0.091   &    0.001   &    0.000   &    0.000   &    0.092   \\
    50     &    0.119   &    0.002   &    0.000   &    0.001   &    0.121   \\
   100     &    0.095   &    0.007   &   -0.001   &    0.001   &    0.102   \\
   150     &    0.057   &    0.011   &   -0.002   &    0.002   &    0.068   \\
   200     &    0.025   &    0.014   &   -0.003   &    0.003   &    0.038   \\
   300     &   -0.018   &    0.018   &   -0.006   &    0.003   &   -0.003   \\
 \hline 
 \\ \multicolumn{6}{c}{$^3F_3$}  \\
    10     &    0.009   &    0.000   &    0.000   &    0.000   &    0.009   \\
    25     &    0.044   &    0.000   &    0.000   &    0.000   &    0.044   \\
    50     &    0.093   &    0.000   &    0.000   &    0.000   &    0.093   \\
   100     &    0.141   &    0.001   &    0.000   &    0.000   &    0.142   \\
   150     &    0.159   &    0.003   &    0.000   &    0.000   &    0.163   \\
   200     &    0.166   &    0.005   &    0.000   &    0.001   &    0.171   \\
   300     &    0.165   &    0.010   &   -0.001   &    0.002   &    0.177   \\
 \hline 
 \\ \multicolumn{6}{c}{$^1G_4$}  \\
    25     &   -0.009   &    0.000   &    0.000   &    0.000   &   -0.009   \\
    50     &   -0.023   &    0.000   &    0.000   &    0.000   &   -0.023   \\
   100     &   -0.032   &    0.001   &    0.000   &    0.000   &   -0.032   \\
   150     &   -0.029   &    0.002   &    0.000   &    0.000   &   -0.027   \\
   200     &   -0.022   &    0.003   &    0.000   &    0.000   &   -0.019   \\
   300     &   -0.005   &    0.006   &    0.000   &    0.000   &    0.001   \\
 \hline 
 \\ \multicolumn{6}{c}{$^3F_4$}  \\
    25     &   -0.004   &    0.000   &    0.000   &    0.000   &   -0.004   \\
    50     &   -0.015   &    0.000   &    0.000   &    0.000   &   -0.015   \\
   100     &   -0.037   &    0.000   &    0.000   &    0.000   &   -0.037   \\
   150     &   -0.053   &    0.000   &    0.000   &    0.000   &   -0.053   \\
   200     &   -0.064   &    0.000   &    0.000   &   -0.001   &   -0.065   \\
   300     &   -0.078   &   -0.001   &    0.000   &   -0.001   &   -0.080   \\
 \hline 
 \\ \multicolumn{6}{c}{$^3H_4$}  \\
    50     &   -0.007   &    0.000   &    0.000   &    0.000   &   -0.007   \\
   100     &   -0.019   &    0.000   &    0.000   &    0.000   &   -0.019   \\
   150     &   -0.028   &    0.000   &    0.000   &    0.000   &   -0.029   \\
   200     &   -0.035   &    0.000   &    0.000   &    0.000   &   -0.036   \\
   300     &   -0.044   &   -0.001   &    0.000   &    0.000   &   -0.045   \\
 \hline 
 \\ \multicolumn{6}{c}{$\epsilon_4$}  \\
    25     &    0.012   &    0.000   &    0.000   &    0.000   &    0.012   \\
    50     &    0.033   &    0.000   &    0.000   &    0.000   &    0.033   \\
   100     &    0.058   &    0.000   &    0.000   &    0.000   &    0.059   \\
   150     &    0.067   &    0.001   &    0.000   &    0.000   &    0.068   \\
   200     &    0.068   &    0.001   &    0.000   &    0.000   &    0.070   \\
   300     &    0.062   &    0.002   &    0.000   &    0.000   &    0.065   \\
\end{tabular}
\end{table}

\begin{table}
\caption{CIB phase shift differences (in degrees) as defined in Eq.\
(9) 
for the various $2\pi$-exchange contributions explained in the text.
$S$ denotes stretched-box diagrams, $B$ box, and $X$ crossed-box
diagrams.}
\begin{tabular}{rrrrrrrr}
T$_{lab}$ (MeV)&$2\pi NN$-$S$& $2\pi NN$-$X$&$2\pi N\Delta$-$B$ &
 $2\pi N\Delta$-$X$ &
 $2\pi\Delta\Delta$-$B$ &$2\pi\Delta\Delta$-$X$& Total $2\pi$  \\
 \hline 
 \hline 
 \\ \multicolumn{8}{c}{$^1S_0$}  \\
   1 & -0.182 &  0.319 &  0.692 & -0.410 & -0.244 &  0.143 &  0.319 \\
   5 & -0.098 &  0.160 &  0.363 & -0.217 & -0.130 &  0.076 &  0.154 \\
  10 & -0.075 &  0.113 &  0.271 & -0.165 & -0.099 &  0.058 &  0.103 \\
  25 & -0.056 &  0.065 &  0.188 & -0.119 & -0.072 &  0.042 &  0.047 \\
  50 & -0.047 &  0.037 &  0.146 & -0.099 & -0.061 &  0.034 &  0.009 \\
 100 & -0.045 &  0.012 &  0.118 & -0.091 & -0.056 &  0.030 & -0.031 \\
 150 & -0.047 & -0.003 &  0.109 & -0.093 & -0.058 &  0.030 & -0.061 \\
 200 & -0.052 & -0.015 &  0.107 & -0.100 & -0.062 &  0.031 & -0.091 \\
 300 & -0.067 & -0.041 &  0.112 & -0.126 & -0.078 &  0.037 & -0.163 \\
 \hline 
 \\ \multicolumn{8}{c}{$^3P_0$}  \\
  10 & -0.003 & -0.003 &  0.001 & -0.003 &  0.000 &  0.000 & -0.009 \\
  25 & -0.010 & -0.010 &  0.002 & -0.009 & -0.001 &  0.000 & -0.029 \\
  50 & -0.018 & -0.020 &  0.002 & -0.017 & -0.002 &  0.000 & -0.053 \\
 100 & -0.025 & -0.030 &  0.003 & -0.024 & -0.002 &  0.001 & -0.078 \\
 150 & -0.027 & -0.036 &  0.002 & -0.028 & -0.003 &  0.001 & -0.091 \\
 200 & -0.029 & -0.041 &  0.002 & -0.030 & -0.003 &  0.001 & -0.101 \\
 300 & -0.031 & -0.050 &  0.000 & -0.035 & -0.004 &  0.001 & -0.118 \\
 \hline 
 \\ \multicolumn{8}{c}{$^3P_1$}  \\
  10 &  0.000 &  0.002 &  0.001 & -0.001 &  0.000 &  0.000 &  0.002 \\
  25 & -0.001 &  0.006 &  0.004 & -0.002 &  0.000 &  0.000 &  0.007 \\
  50 & -0.002 &  0.012 &  0.010 & -0.003 & -0.001 &  0.001 &  0.016 \\
 100 & -0.003 &  0.021 &  0.021 & -0.006 & -0.001 &  0.001 &  0.032 \\
 150 & -0.004 &  0.028 &  0.030 & -0.008 & -0.002 &  0.002 &  0.046 \\
 200 & -0.004 &  0.034 &  0.038 & -0.009 & -0.002 &  0.002 &  0.058 \\
 300 & -0.004 &  0.043 &  0.054 & -0.010 & -0.003 &  0.003 &  0.083 \\
 \hline 
 \\ \multicolumn{8}{c}{$^1D_2$}  \\
  50 &  0.000 &  0.003 &  0.002 &  0.000 &  0.000 &  0.000 &  0.005 \\
 100 &  0.000 &  0.009 &  0.007 & -0.001 &  0.000 &  0.000 &  0.015 \\
 150 &  0.000 &  0.014 &  0.013 & -0.002 & -0.001 &  0.001 &  0.025 \\
 200 &  0.000 &  0.017 &  0.019 & -0.003 & -0.001 &  0.001 &  0.034 \\
 300 &  0.000 &  0.020 &  0.031 & -0.004 & -0.002 &  0.002 &  0.047 \\
 \hline 
 \\ \multicolumn{8}{c}{$^3P_2$}  \\
  10 & -0.001 &  0.001 &  0.001 & -0.001 &  0.000 &  0.000 &  0.000 \\
  25 & -0.003 &  0.002 &  0.005 & -0.004 & -0.001 &  0.001 & -0.001 \\
  50 & -0.006 &  0.004 &  0.011 & -0.010 & -0.004 &  0.002 & -0.003 \\
 100 & -0.009 &  0.004 &  0.020 & -0.018 & -0.008 &  0.004 & -0.007 \\
 150 & -0.010 &  0.004 &  0.027 & -0.022 & -0.011 &  0.005 & -0.007 \\
 200 & -0.010 &  0.003 &  0.031 & -0.024 & -0.014 &  0.007 & -0.007 \\
 300 & -0.009 &  0.003 &  0.037 & -0.024 & -0.018 &  0.008 & -0.004 \\
\end{tabular}
\end{table}

\pagebreak

\begin{table}
\caption{CIB contributions to the $^1S_0$ scattering length,
$\Delta a_{CIB}$, 
and effective range, 
$\Delta r_{CIB}$,
from various components of the $NN$ interaction as explained
in the text.}
\begin{tabular}{rrrrrrr}
   & OPE & $2\pi$  & $\pi\rho$ & $\pi\sigma+\pi\omega$ & Total&Empirical  \\
 \hline 
$\Delta a_{CIB}$ (fm) &
 3.243 & 0.360 & -0.383 & 1.426 & 4.646 & $5.7\pm 0.5$
\\
$\Delta r_{CIB}$ (fm) &
 0.099 & 0.002 & -0.006 & 0.020 & 0.115 & $0.05\pm 0.13$
\\
\end{tabular}
\end{table}

\vspace*{2cm}

\begin{table}
\caption{CIB contributions to the $^1S_0$ scattering length,
$\Delta a_{CIB}$, and effective range, $\Delta r_{CIB}$,
for the various parts of the $2\pi$-exchange model
as explained in the text.
$S$ denotes stretched-box diagrams, $B$ box, and $X$ crossed-box
diagrams.}
\begin{tabular}{rrrrrrrr}
   &$2\pi NN$-$S$& $2\pi NN$-$X$&$2\pi N\Delta$-$B$ &
 $2\pi N\Delta$-$X$ &
 $2\pi\Delta\Delta$-$B$ &$2\pi\Delta\Delta$-$X$& Total $2\pi$  \\
 \hline 
$\Delta a_{CIB}$ (fm) &
 -0.196 & 0.351 & 0.787 & -0.470 & -0.272 & 0.159 & 0.360
\\
$\Delta r_{CIB}$ (fm) &
-0.003 & 0.003 & 0.010 & -0.006 & -0.004 & 0.002 & 0.002
\\
\end{tabular}
\end{table}

\pagebreak

\begin{figure}
\caption{One-pion-exchange (OPE) contribution to 
(a) $pp$ and (b) $np$ scattering.}
\end{figure}

\begin{figure}
\caption{Irreducible 2$\pi$-exchange diagrams with $NN$ intermediate
states for (a) $pp$ and (b) $np$ scattering.}
\end{figure}

\begin{figure}
\caption{2$\pi$-exchange contributions with $N\Delta$ intermediate
states to (a) $pp$ and (b) $np$ scattering.}
\end{figure}

\begin{figure}
\caption{2$\pi$-exchange contributions with $\Delta\Delta$ intermediate
states to (a) $pp$ and (b) $np$ scattering.}
\end{figure}

\begin{figure}
\caption{CIB phase shift differences $\Delta\delta^{LSJ}_{CIB}$ (in degrees)
as defined in Eq.~(9) for laboratory kinetic energies $T_{lab}$
below 300 MeV and partial waves with total angular momentum
$J\leq 2$.
The CIB effects due to OPE, the entire $2\pi$ model,
$\pi\rho$ exchanges, and $(\pi\sigma+\pi\omega)$ contributions
are shown by the dashed, dash-dot, dash-triple-dot, and dotted
curves, respectively.
The solid curve is the sum of all CIB effects.
(See text for further explanations.)}
\end{figure}

\begin{figure}
\caption{Similar to Fig.~5, but here the individual contributions
from the 2$\pi$ model are shown.
The CIB effects due to $2\pi NN$, $2\pi N\Delta$-$B$, $2\pi N\Delta$-$X$,
and $2\pi \Delta\Delta$
are shown by the dashed, dash-dot, dash-triple-dot, and dotted
and $2\pi \Delta\Delta$
are shown by the dashed, dash-dot, dash-triple-dot, and dotted
curves, respectively.
The solid curve is the sum of all CIB effects due to the
exchange of two pions.
(See text and caption of Table II for further explanations.)}
\end{figure}


\begin{thebibliography}{99}
\bibitem{foot}
The relationship between $g_\pi$ and $f_{\pi NN}$ [the latter being used
in the Lagrangian Eq.~(7)] is
\begin{displaymath}
g^2_\pi = \left( \frac{2M_N}{m_{\pi^\pm}} \right)^2
 f^2_{\pi NN} = 181.027 f^2_{\pi NN} \; .
\end{displaymath}
Following the Bonn model~\cite{MHE87}, 
we also assume a fixed relationship between 
$f_{\pi N\Delta}$ [used in the Lagrangian Eq.~(8)] and $f_{\pi NN}$,
namely,
\begin{displaymath}
f^2_{\pi N\Delta} = \frac{72}{25} f^2_{\pi NN} \; .
\end{displaymath}
\end{thebibliography}
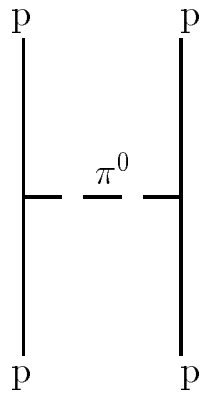
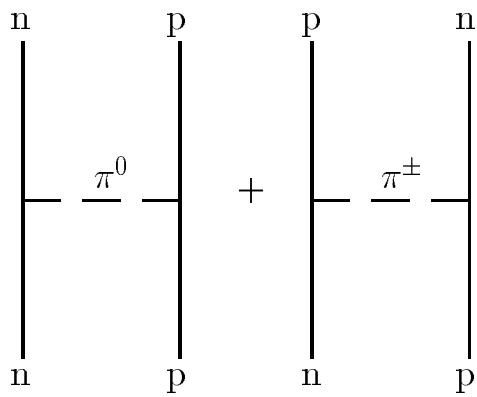
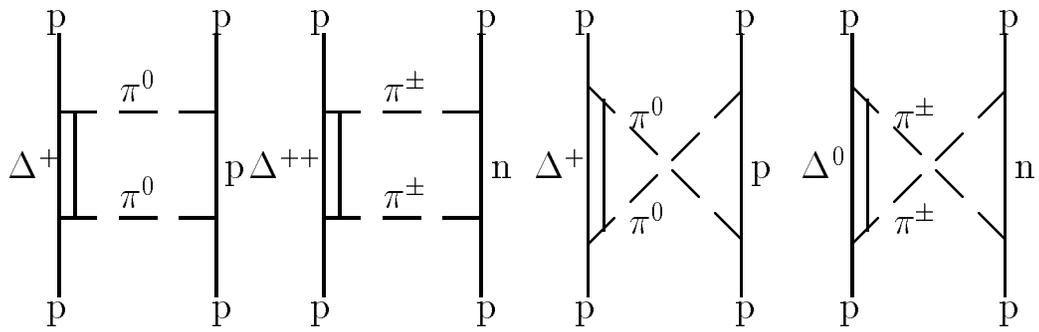
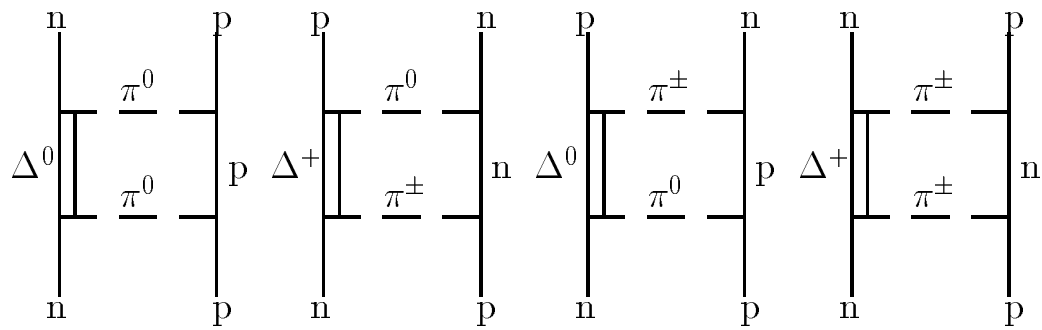
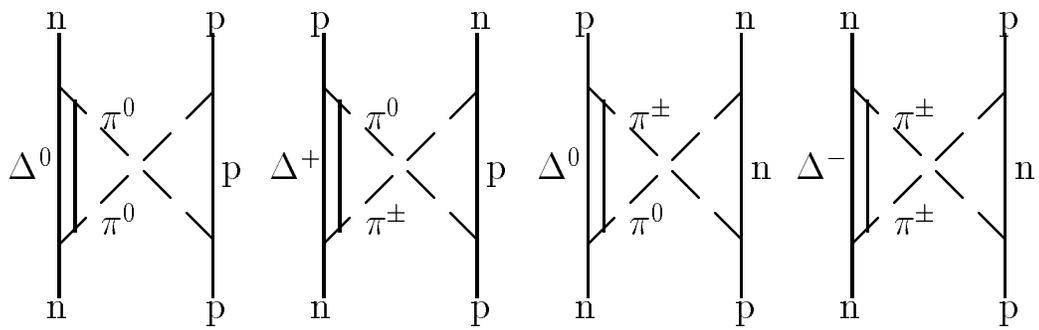
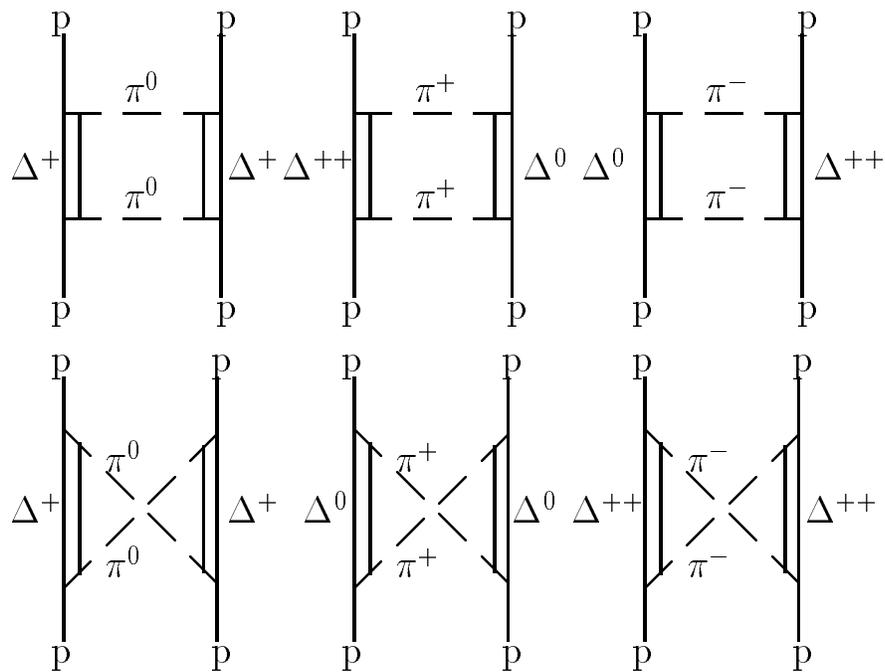
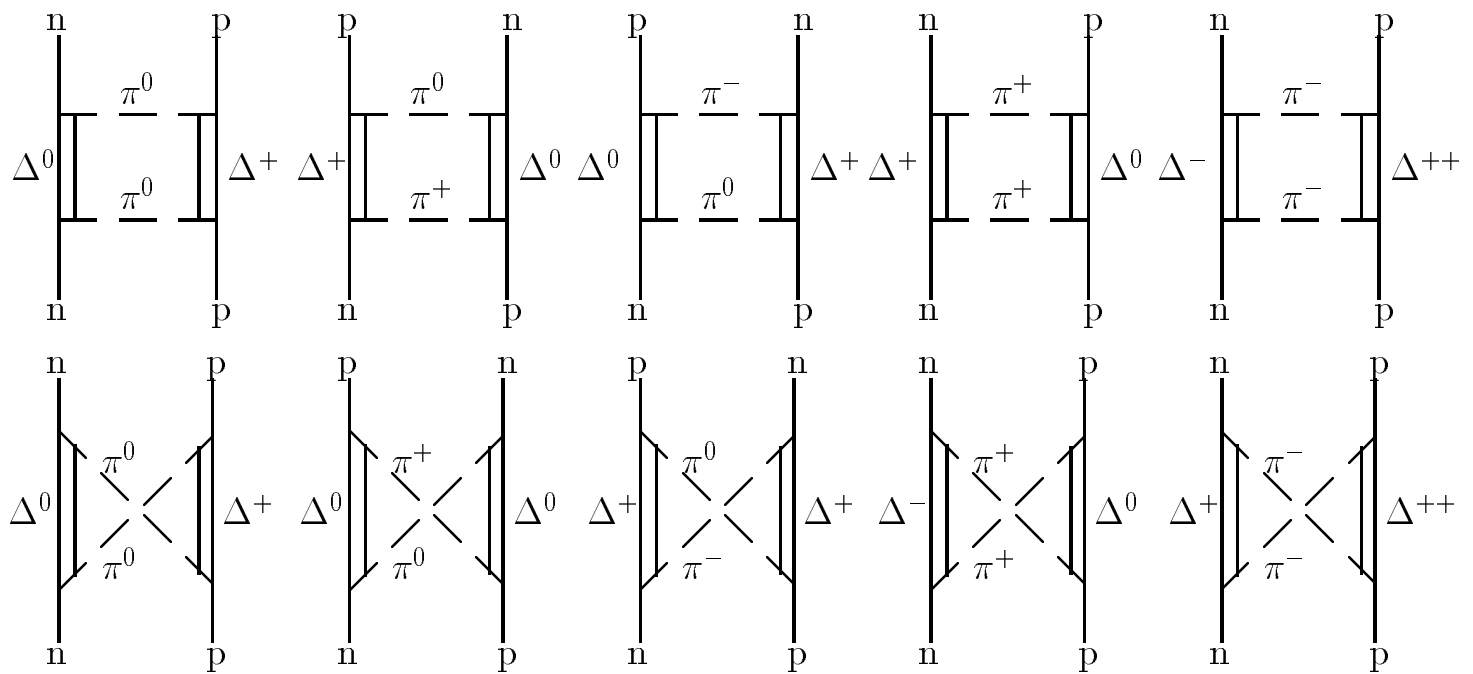
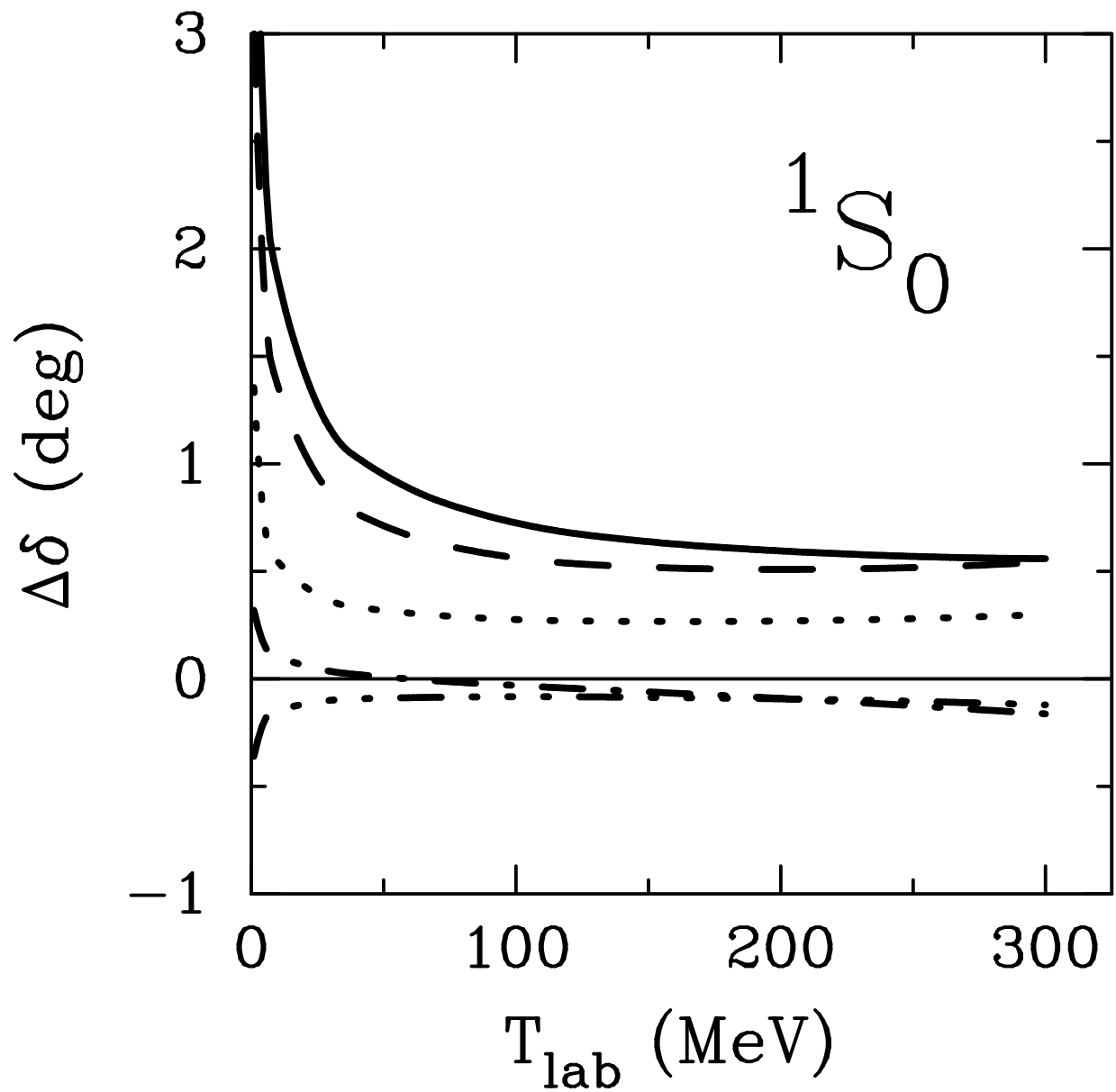
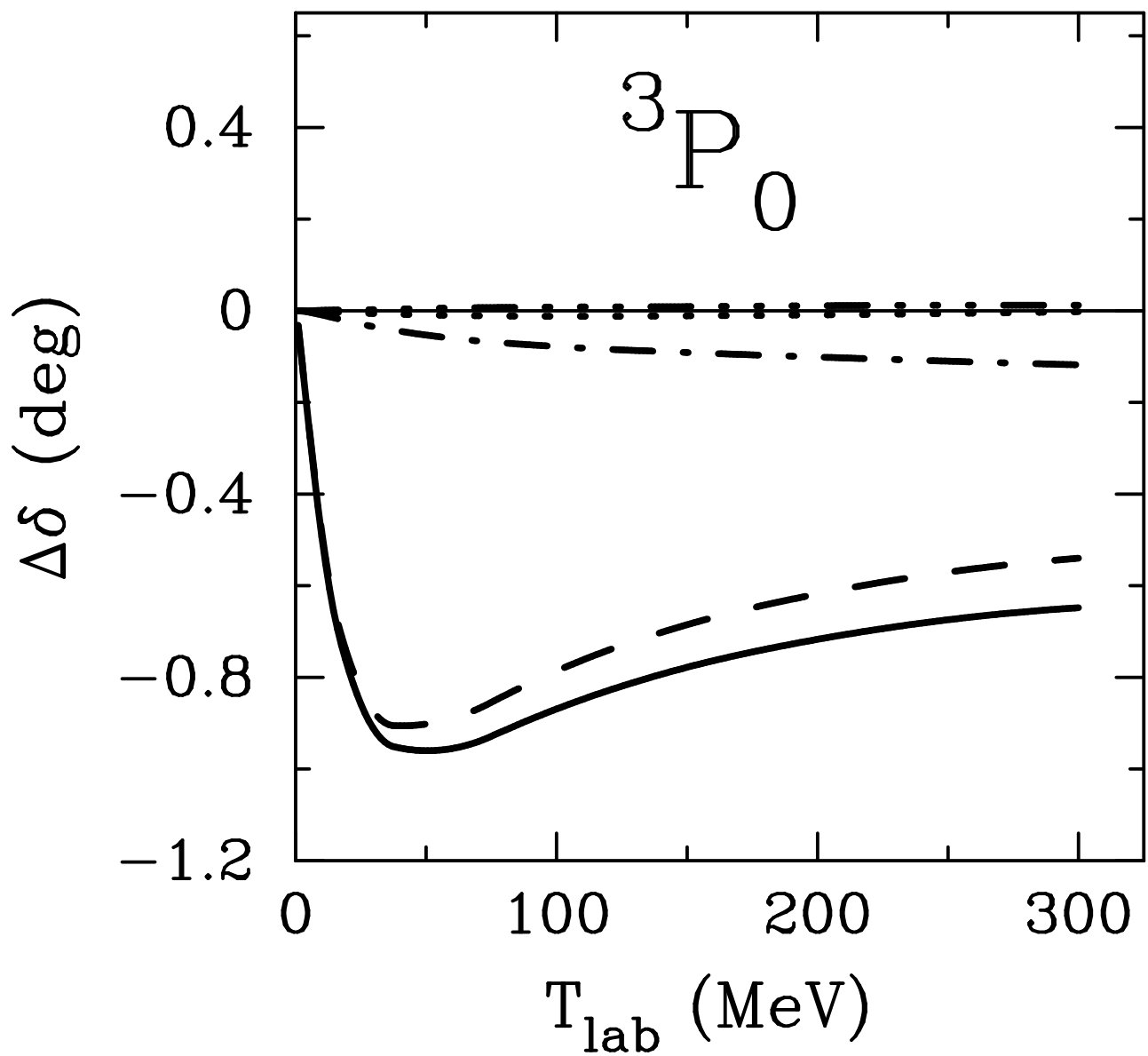
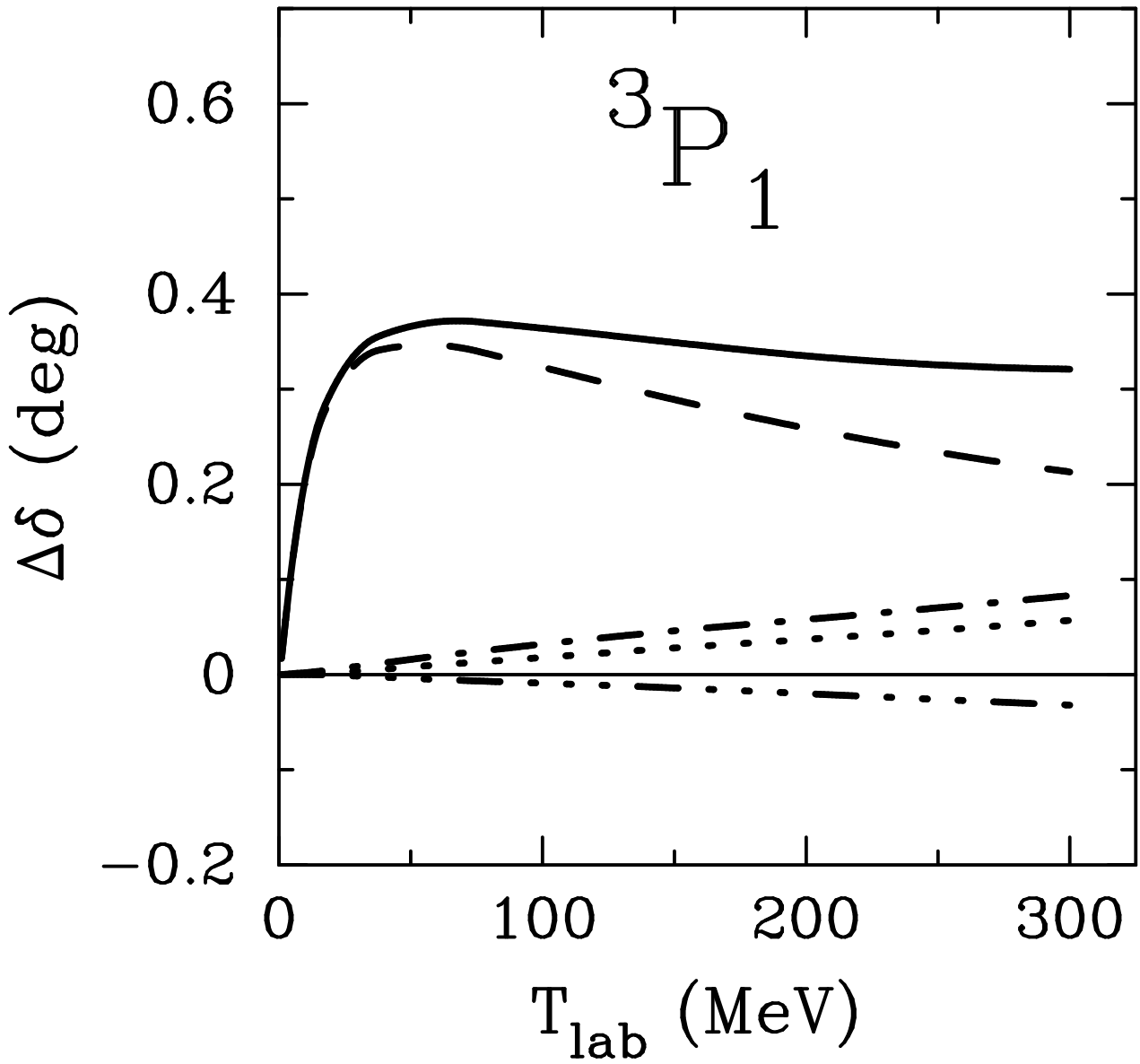
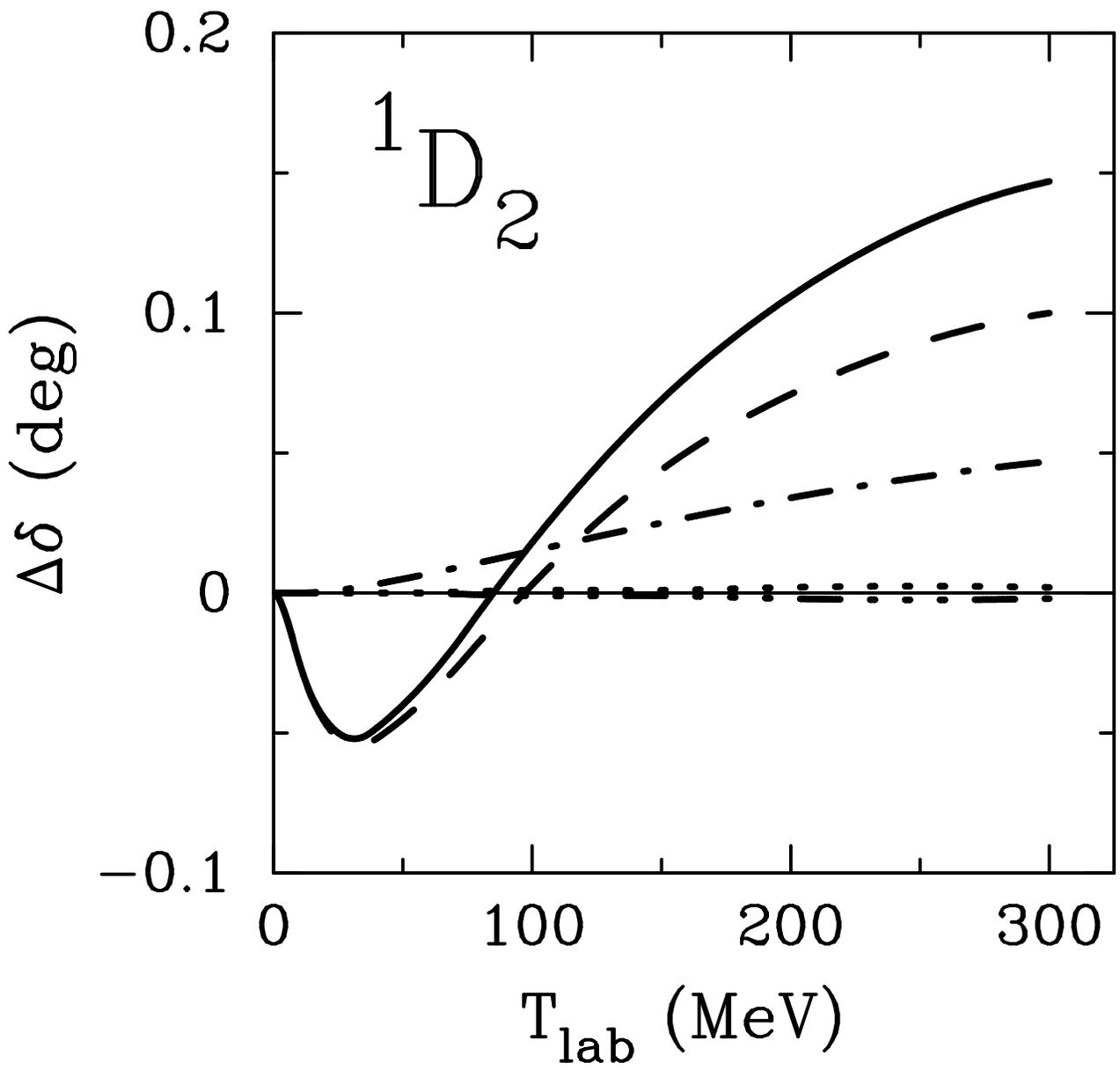
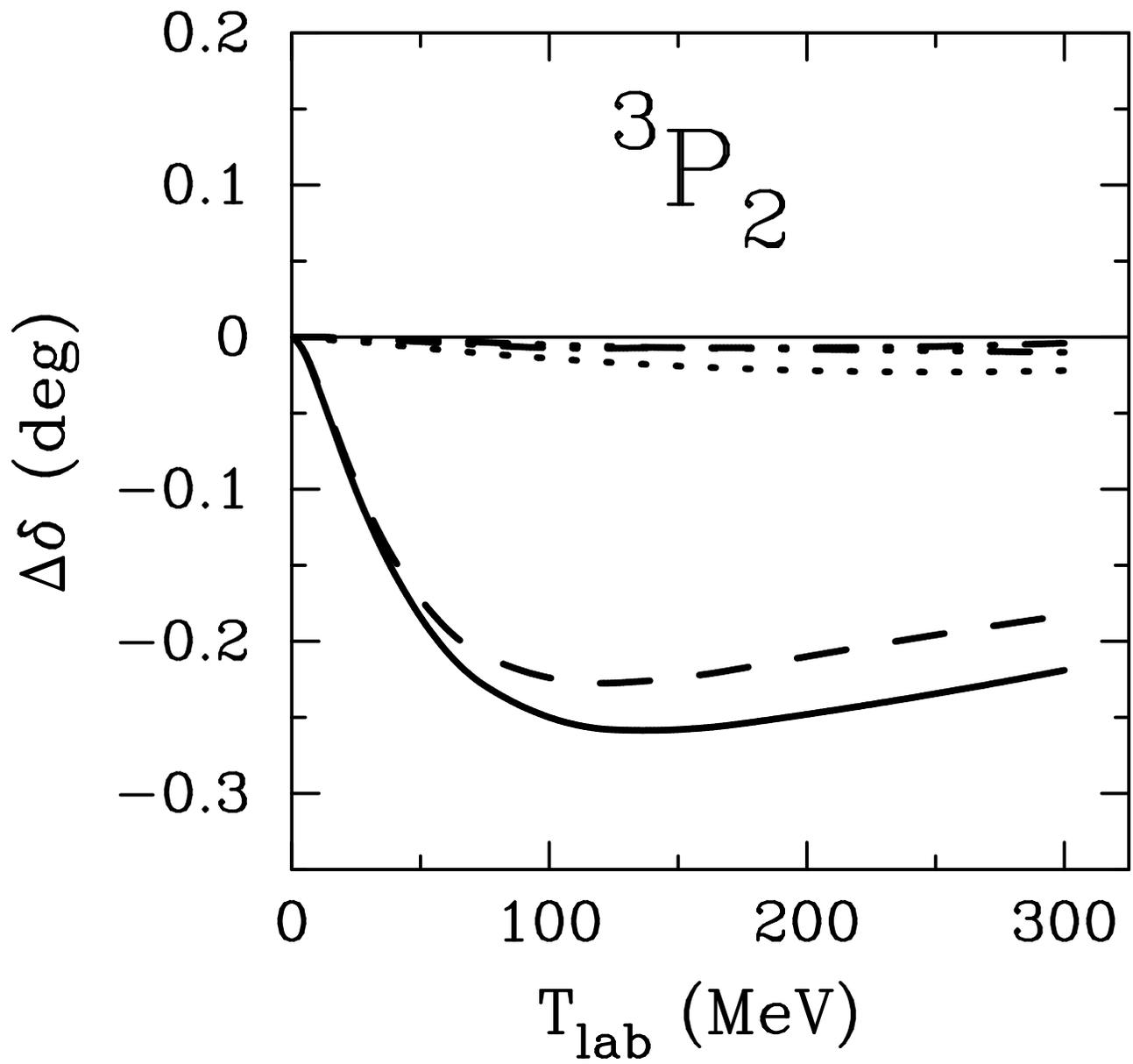
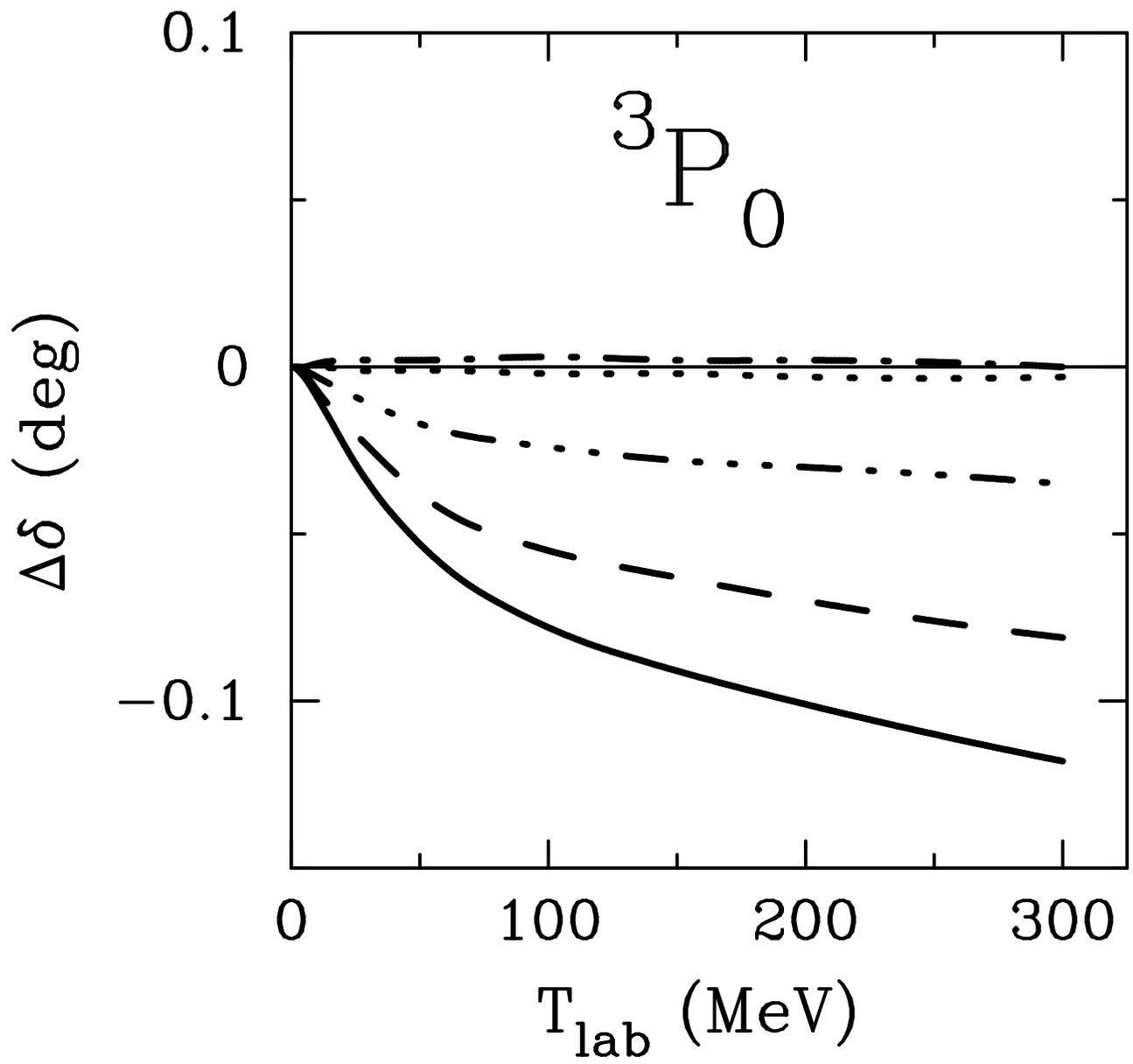
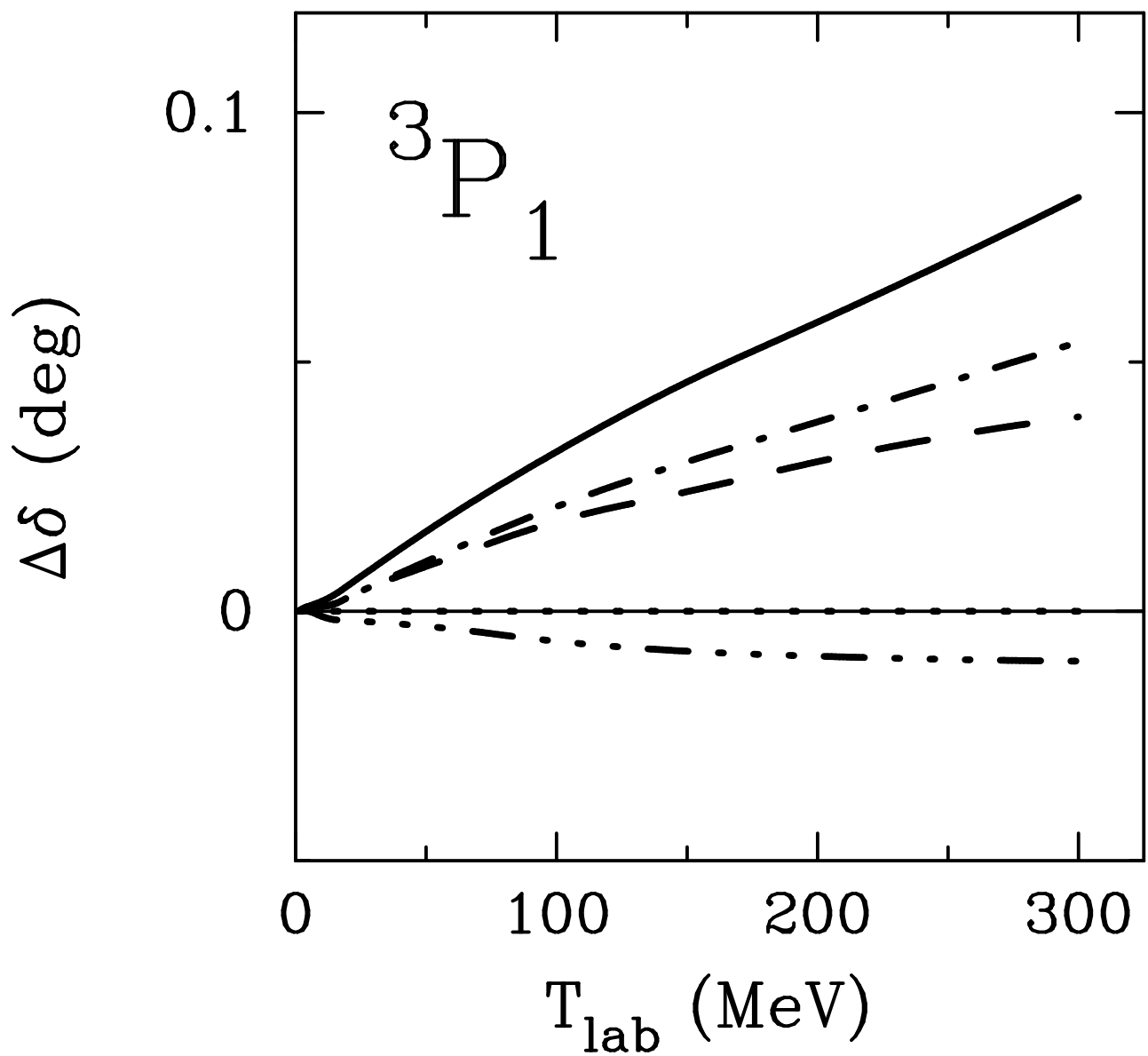
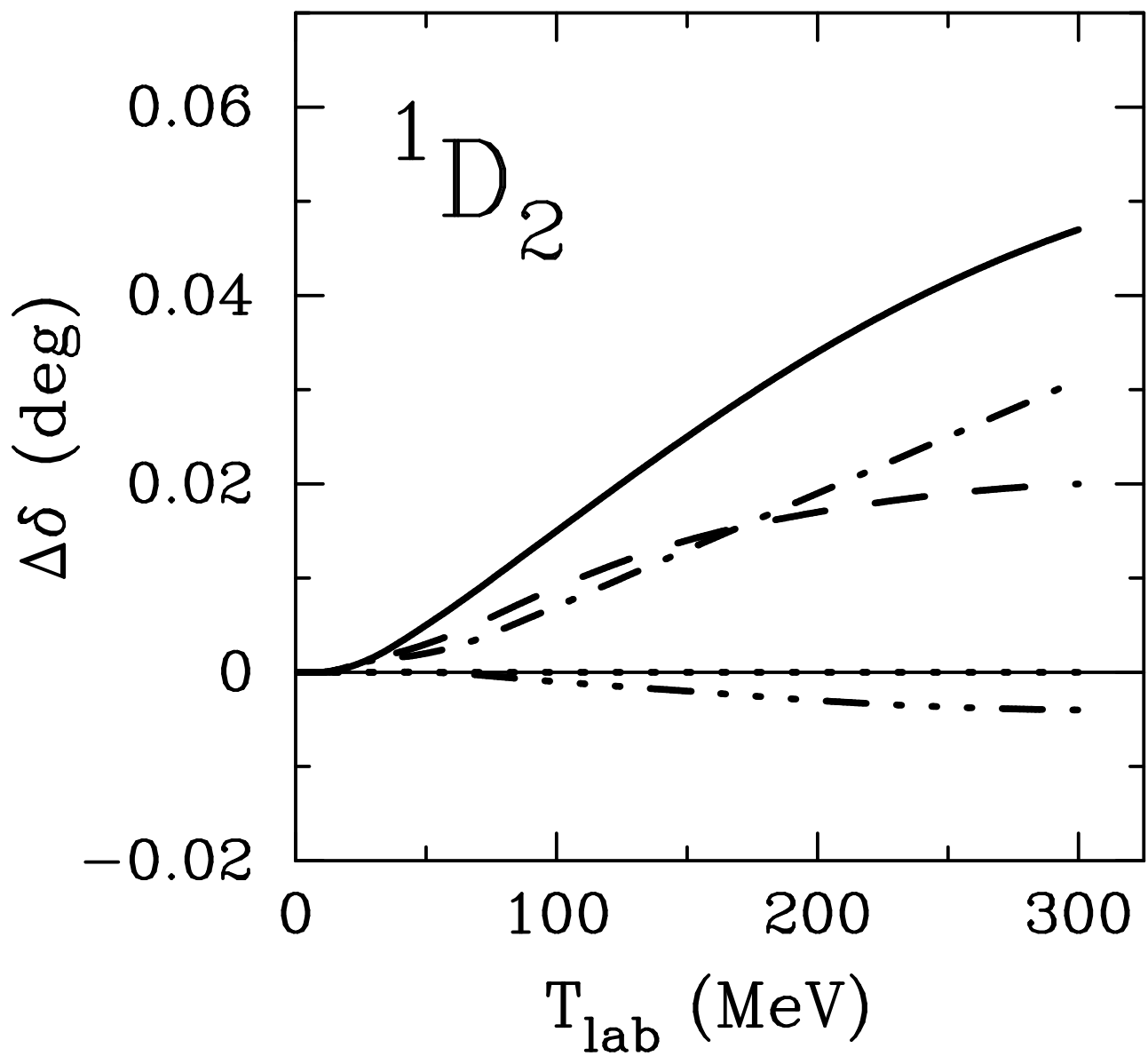
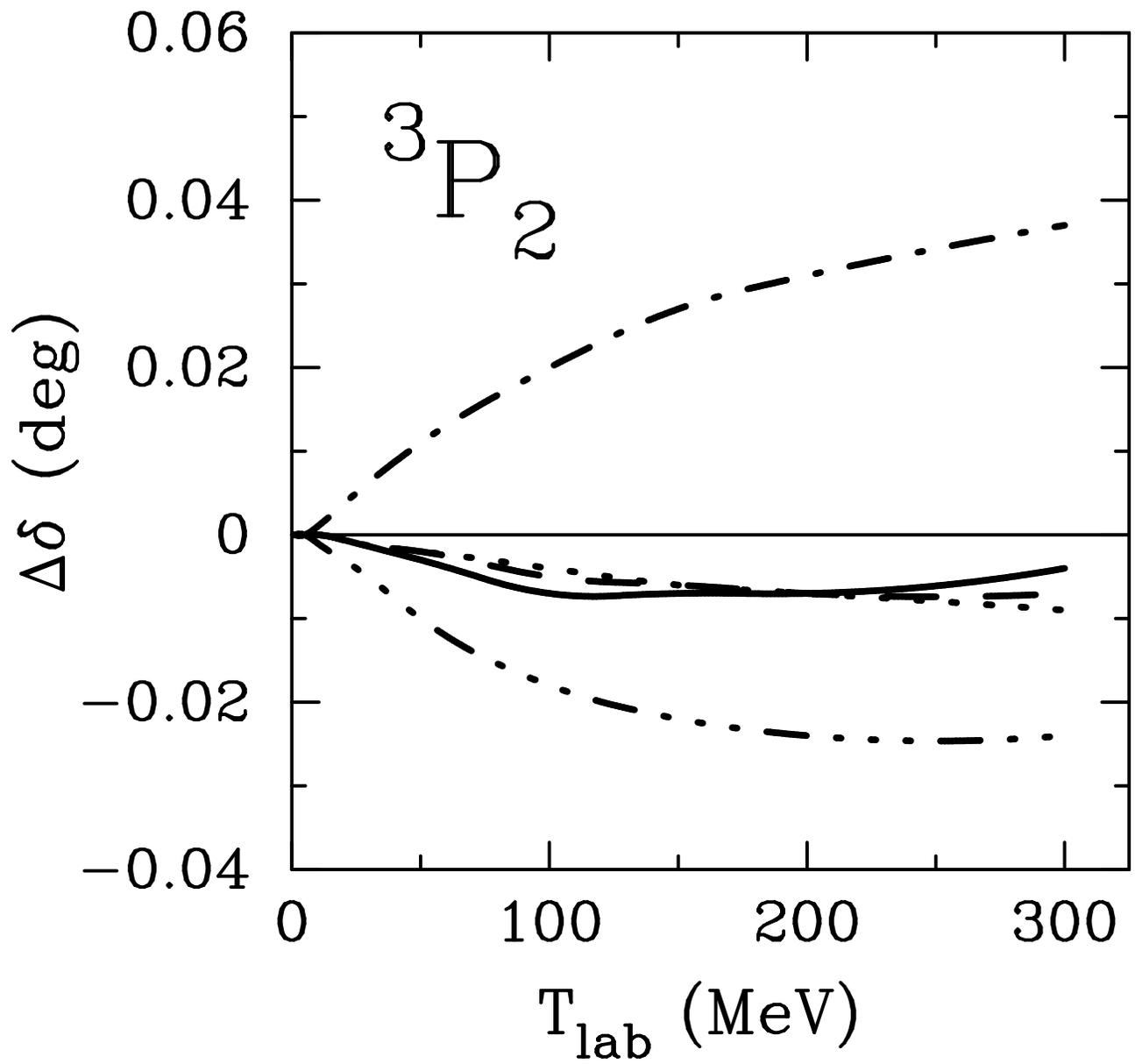
\end{document}